\documentclass{iau}

\title[Magnetic fields in our Milky Way Galaxy and nearby galaxies] 
{Magnetic fields in our Milky Way Galaxy and nearby galaxies
} 

\author[J. L. Han]  
{JinLin Han 
}

\affiliation{
National Astronomical Observatories, Chinese Academy of Sciences,
\\ DaTun Road 20A, ChaoYang District, Beijing 100012, China
}

\pubyear{2013}
\volume{294}  
\pagerange{213--224}
\jname{\mbox{Solar and Astrophysical Dynamos and Magnetic Activity}}
\editors{A.G. Kosovichev, E.M. de Gouveia Dal Pino, \&  Y.Yan, eds.}
\begin{document}

\maketitle

\begin{abstract}
Magnetic fields in our Galaxy and nearby galaxies have been revealed
by starlight polarization, polarized emission from dust grains and
clouds at millimeter and submillimeter wavelength, the Zeeman effect
of spectral lines or maser lines from clouds or clumps, diffuse radio
synchrotron emission from relativistic electrons in interstellar
magnetic fields, and the Faraday rotation of background radio sources
as well as pulsars for our Milky Way. It is easy to get a global
structure for magnetic fields in nearby galaxies, while we have
observed many details of magnetic fields in our Milky Way, especially
by using pulsar rotation measure data. In general, magnetic fields in
spiral galaxies probably have a large-scale structure. The fields
follow the spiral arms with or without the field direction
reversals. In the halo of spiral galaxies magnetic fields exist and
probably also have a large-scale structure as toroidal and poloidal
fields, but seem to be slightly weaker than those in the disk. In the
central region of some galaxies, poloidal fields have been detected as
vertical components. Magnetic field directions in galaxies seem to
have been preserved during cloud formation and star formation,
from large-scale diffuse interstellar medium to molecular clouds and
then to the cloud cores in star formation regions or clumps for the
maser spots. Magnetic fields in galaxies are passive to
dynamics.
\end{abstract}

\firstsection 

\section{Introduction}

Magnetic fields are frozen in the interstellar medium. Their effects are
easily detectable together with the interstellar gas. Elliptical
galaxies do not have much interstellar gas, and the magnetic fields of
elliptical galaxies appear only in the jets originating from the
central blackhole. Therefore, when talking about magnetic fields in
nearby galaxies, we most probably mean those in spiral galaxies.

Magnetic fields exist on all scales in the interstellar medium of
spiral galaxies, from the stellar scale of AU size, especially in the
star formation regions, to the galactic scale of tens of kpc along the
spiral arms.

There are many probes for interstellar magnetic fields.  The most
widely used are the starlight polarization, polarized emission from
dust grains and clouds at millimeter and submillimeter wavelengths,
the Zeeman effect of spectral lines or maser lines from clouds or
clumps, diffuse radio synchrotron emission from relativistic electrons
in interstellar magnetic fields, and the Faraday rotation of
background radio sources as well as pulsars for our Milky Way. The
first three are related to magnetic fields in clouds, and the later
two are related to the fields in the diffuse medium. All these methods
have been used to detect the magnetic fields in the Milky Way and
nearby galaxies.  Note that any of these types of magnetic probes can
only reveal information of one of the three dimensional components of
magnetic fields, except for Zeeman splitting.  Therefore, when we look
at the observational results, it is important to understand in which
part of a galaxy the magnetic fields have been measured, and how the
results are related to the magnetic field properties in a galaxy.

In analogy to the situation for Solar and stellar magnetic field studies,
we have observed many more details of magnetic fields in our
Milky Way galaxy than those in nearby galaxies. We often see a lot of
``trees'' for magnetic fields in our own Milky Way, while we
see ``the forest'' in nearby galaxies. I will
first review and comment on the observational results of magnetic fields
in nearby galaxies, and then review the main results of the magnetic
structure in our Milky Way. To save space for more context and references,
no figures are included in this review. Readers are encouraged to
look at PPT file at web-page: http://zmtt.bao.ac.cn/hjl/IAU294Han.ppt

\section{Magnetic fields in nearby galaxies}

Nearby galaxies can be spiral galaxies, elliptical galaxies, or
irregular galaxies. Spiral galaxies could be face-on or edge-on or
inclined. Magnetic fields in spiral galaxies in general should have
similar properties to the fields in our Milky Way. Magnetic fields
in nearby galaxies have been detected by using the five types of
probes we mentioned above. Let us review one by one.

\subsection{Optical polarization}

Starlight polarization is caused by the absorption of  dust
grains which are preferentially oriented by interstellar magnetic
fields. Optical polarimetry has been used for the images of nearby
galaxies since 1970 (eg. \cite{mf70} for the Large Magellanic Cloud
(LMC); \cite{sww87} for M51; \cite{srwt91} for NGC 1068).
Polarization images show that the polarization vectors and hence the
magnetic fields are oriented along the spiral arms.

Now the optical polarimetry and the CCD camera have been well
developed. Many new instruments have been constructed for near or far
infrared polarization images (e.g. \cite{mdc+12}; \cite{cppt12}),
which have already been used for observations of detailed magnetic
fields of clouds by using the polarized light of stars behind the
clouds (e.g. \cite{mpc12}). Probably a new era is coming to use these
instruments for revealing the magnetic fields in {\it inclined}
galaxies with enough dust grains, showing magnetic field details in
different parts of a galaxy, much more sensitive than previously has
been done by e.g. Scarrott and his team (see the review by
\cite{sca96}). But at present not many results have been
published. Observations for {\it face-on} galaxies probably cannot
obtain images with much significant polarization (e.g. \cite{pc12}).

\subsection{Polarized emission of clouds and dust grains}

Clouds and dust grains have thermal emission with a temperature in a
range of a few tens to a few hundred K, and the radiation is peaked at
the mm or submm wavelength or far-infrared. The polarized emission
from the clouds or dust grains can show the magnetic field orientations in
the clouds perpendicular to the line of sight. The technology for
polarimetry at these wavelengths has developed very well in the last
decades (e.g. \cite{hdd+00}). Excellent observations have been
achieved in Galactic objects and the Galactic plane
(e.g. \cite{lgk+06}; \cite{bmd+11}). However, the sensitivity and
resolution are not good enough for objects in external nearby galaxies.  Up
to now very few observations have been carried out. The best are
observations of 6 giant molecular clouds in M33 by using the SMA
(\cite{lh11}). The fields shown by the polarization vectors in these 6
giant molecular clouds are aligned with the spiral arms, suggesting
that the large-scale field in M33 anchors the clouds. In the near future,
polarization images of nearby galaxies at mm or submm wavelength
can probably be obtained by large telescopes, such as ALMA or larger
single dish submm telescopes.

\subsection{Maser emission}

Maser emission comes from the dense core or clumps in a cloud. The
strong nearby pump sources can be bright stars or galactic nuclei.
The line emission or absorption from the clouds or clumps permeated by
magnetic fields shows Zeeman splitting. The separation
of split lines measures the field strength of the magnetic
component parallel to the line of sight, and the sense change of
circular polarization of split lines indicates the direction of
the field component along the line of sight.

Because of the limited sensitivity, only megamasers in nearby galaxies
near their galactic nuclei have been detected. The Zeeman splitting of the
megamasers have been detected from 4 and 11 ultra luminous infrared galaxies
(\cite{rqh08}, \cite{mh13}), which indicate the magnetic fields in the star
formation regions near the galactic nuclei have a strength of a few mG
to 30 mG, similar to the magnetic fields in star formation regions in
our Milky Way (see Table 1 in \cite{hz07}). Though it is not
possible now, in the future, one could use a large telescope array,
e.g. SKA, to observe the Zeeman splitting of many normal maser regions
in the disk of nearby galaxies and hence to outline the global field
directions (not merely orientations), and to check if the magnetic fields in
molecular clouds and star formation regions are correlated well with
the global spiral and magnetic structures (\cite{hz07}).

\subsection{Diffuse radio polarized emission}

Synchrotron emission of relativistic electrons gyrating magnetic
fields certainly gives information about magnetic fields. The total
radiation intensity is related the total strength of fields, and the
polarized intensity and polarization angles are related to the
strength fraction and the field orientation (not directions) in the
sky plane of ordered fields or anisotropic fields. Synchrotron
emission produced in random magnetic fields does not show
polarization because the polarized emission from all small volume
cells within a telescope beam are summed together and hence
depolarized.

The first polarized emission images of spiral galaxies, M51, M81 and
M31, were observed using WSRT (\cite{mvb72}, \cite{ssd76}). The images
show the E-vector perpendicular to the arms and optical vectors so
that the ordered magnetic fields are indicated to follow the spiral
arms. The German MPIfR group led by Prof. R. Wielebinski has been
dedicated to the polarization observations of many nearby galaxies at
many bands for more than two decades by using the VLA and Effelsberg
telescopes (e.g. \cite{bbw78}, \cite{kbh89}, \cite{nkbw91},
\cite{bh96}, \cite{bse+02}, \cite{wk93}). I visited the group and was
involved in the studies of magnetic fields in NGC 2997
(\cite{hbe+99}), a beautiful grand-design spiral galaxy as always seen
in text books.  For spiral galaxies, the observed polarization
vectors, and hence the B-vectors after being rotated by $90^{\circ}$
at a short wavelength, show that the ordered fields follow the spiral
arms. Strong polarized emission is detected in interarm regions
(\cite{bh96}).  The most impressive is that the fields follow the arm
even when an arm is locally distorted (e.g. \cite{hbe+99}) or
dynamically distorted in the interaction pairs (e.g.
\cite{fbs+11}). This means that magnetic fields are dynamically
passive, which has been confirmed by polarization observations of bar
galaxies (e.g. \cite{bse+02}). Polarized emission has also been 
detected in the halos of some edge-on galaxies (e.g. \cite{hbd91}),
indicating an X-shaped field structure as poloidal-like (e.g. NGC
4631) or simply parallel to the mid-plane as being toroidal-like.

A Polish group has joined the efforts for magnetic fields in nearby
galaxies and often cooperated with the German group. They have observed
many different types of galaxies, e.g. irregular
galaxies (e.g. \cite{cbk+00}, \cite{ckb+03}),
merging or interacting galaxies (e.g. \cite{sub+01}, \cite{cb04}),
ring galaxies (\cite{cb08}), flocculent galaxies (\cite{sbub02}). The
magnetic fields are revealed to always have a good patten, but
again are passive to dynamics. In other words, magnetic fields in galaxies
cannot affect dynamics, but dynamics can affect magnetic fields.

Notice that the polarized emission could be produced by either
large-scale ordered coherent magnetic fields or anisotropic
random fields. To judge if the magnetic fields in the nearby spiral
galaxies have large-scale coherent directions, a rotation measure
(RM) map is necessary. The images of polarization angles in two or more
wavelengths have often been used to derive rotation measure
maps and study the magnetic field structure (e.g. \cite{bbh03},
\cite{fbs+11}). Polarization maps with marginal significance
(3-5$\sigma$) in most areas can produce
an RM map but only significant in some area, not everywhere
(e.g. \cite{hbe+99}).
Note also that the polarized emission of nearby galaxies
has different depth limits at different wavelengths. If
observations are made at high frequencies, e.g. 5GHz or higher, the
polarized emission from all layers in the disk for the whole thickness
is transparent and has a peak near the galactic plane. The emission
can be added together without much Faraday rotation. If the
observations are made at longer wavelengths, e.g. 20~cm or longer, the
polarized emission from the far layers suffers from severe and various
Faraday rotations and hence is depolarized when emission from these
far layers is added together. Only polarized emission from nearer
shallow layers of the disk can be observed at a long
wavelength band. The observable layer depth depends on the wavelength
and the inclination angle of a galactic disk.

The global variation of rotation measures derived from 
polarization maps at two or more almost transparent frequencies can
show the magnetic field configuration of the nearer layers up to a
half thickness of the disk. Polarization angle maps of many
channels of a very long wavelength band therefore can reveal the
field structure in the shallow layers but not the global field
configuration in a galactic disk of whole thickness.

In recent years we see the great developments on polarimetry observations
of many channels in wide bands, i.e. the spectro-polarimetry. When
polarization angles of many frequency channels have been measured, one
can get the rotation measure maps or perform  rotation measure
synthesis (\cite{bd05}). New observations made with the EVLA for 35 edge-on
galaxies (\cite{ibb+12a,ibb+12b}) and with  Westbork for 21 nearby
galaxies (\cite{hbe09}, \cite{bhb10}) have achieved the rotation
measure maps, from which magnetic field structures in the galactic
disk, up to a half thickness, have been derived. Many small-scale field
structures emerge in the RM map (e.g. \cite{hea12}), in addition to the
general large-scale RM distribution.

\subsection{Rotation measures of background polarized sources}

If observational sensitivity is good enough, many background radio
sources behind a large galaxy can be detected. If they are polarized,
and their RMs can be measured, then the RMs can be used to diagnose the
magnetic fields in the galaxy for the whole thickness of the disk.
Magnetic fields of an odd mode or even mode cause different spacial RM
distributions of background sources (\cite{hbb98}). The more RMs of
background sources are observed, the better the field configuration in
the galaxy can be diagnosed. In this aspect, our Milky Way galaxy has
the major advantage that all the polarized radio sources in the sky can be
used as the probes for the magnetic field configuration in both the
disk and the halo (see next section). For external galaxies, at present
the sensitivity for radio polarization observations is very limited,
so only a limited number of radio sources behind a few large objects have
been observed.

The first one is M31 (\cite{hbb98}). In a field of view of a few
square degrees, the rotation measures of 21 polarized radio sources
have been observed.  Using these rotation measures, together with the
rotation measures derived from the diffuse polarized radio emission
from the galactic disk at 6cm and 11cm, Han et
al. (1998)\nocite{hbb98} showed that the magnetic fields in M31 may
have an even mode field configurations.

Two of the largest objects are in the southern sky. One is the LMC.
By using ATCA, Gaensler et al. (2005)\nocite{ghs+05} have observed RMs
of 240 sources behind or around the LMC, and figured out the magnetic
field configuration in the LMC as being axisymmetric fields. Another one
is Cen A. Feain et al. (2009, 2011)\nocite{fce+11,fem+09} have used the
ATCA to observe it for 1200 hours, and got 281 RMs behind or around
Cen A. By comparing the 121 RMs behind the lobes and 160 sources
outside the lobes, they found that the lobes contribute very small
RMs of a few rad~m$^{-2}$, but the lobes cause RM fluctuations of
20-30 rad~m$^{-2}$. Note that this is the best magnetic field 
measurements for the jets from an elliptical galaxy.

In the future, because of excellent sensitivity, the SKA may observe RMs
of many background sources for a few hundred galaxies (\cite{gae06}),
and can probe magnetic field configurations in these galaxies.

\subsection{Summaries for external galaxies}

Many kinds of observations for magnetic fields in nearby galaxies have
been made. Optical starlight polarization can be used to trace
magnetic field orientations in gas-rich properly-inclined
galaxies. The polarization observations and the Zeeman splitting
observations for magnetic fields in clouds or cloud cores or clumps in
nearby galaxies are still very premature because of currently poor
sensitivity and resolution. The most extensive measurements are radio
continuum polarization observations. The polarization vectors show 
that magnetic field orientations follow the spiral arms and are evidently
dynamically passive. The rotation measure maps of polarized
diffuse emission can be used to probe the magnetic field directions in
the galactic disk of the nearer half thickness. Rotation measure
synthesis for the multichannel polarization observations in a very
wide band has been used to diagnose the magnetic configurations. The
rotation measures of background radio sources behind a galaxy can be
used to probe the magnetic configurations in a galactic disk of the
full thickness.

\section{Magnetic fields in our Milky Way galaxy}

The Milky Way can be divided into three parts: the central region, the
halo and the disk. Many details of magnetic fields in our Milky Way
have been revealed by using the five probes mentioned above. Because
of our location at the disk edge of our Galaxy, the best probes for
the {\it large-scale structure} of the Galactic magnetic fields have
to be able to detect the magnetic component parallel to the line of
sight. For this reason, three probes, which show the orientation
of the transverse field component: starlight polarization, the
polarized emission of dust grains and clouds and the diffuse radio
synchrotron emission, are good at revealing magnetic field details
(``trees'') but not at getting the large-scale structure (``the
forest''). The two kinds of excellent probes for the large-scale
magnetic structure are 1) the Zeeman effect of spectral lines or maser
lines from clouds or clumps, if magnetic fields in clouds are closely
related to large-scale magnetic structure, and 2) the Faraday rotation
of extragalactic radio sources and pulsars.

Now I review the observational results as well as new progress for the
measurements of magnetic fields in the three parts of the Milky Way,
by using all kinds of probes. Note that when we try to understand the
properties of magnetic fields in our Milky Way, we have to ``connect''
all measurements in different regions to get the most probable
structure of the Galactic magnetic fields.

\subsection{Magnetic fields in the Galactic central region}

In the central region of a few hundred pc of our Milky Way, the
poloidal magnetic fields are indicated by the polarized radio
filaments, and the toroidal magnetic fields have been detected by
using the polarized emission of the central cloud zone. Recently the
near-infrared polarimetry of stars in the central region shows the
smooth transition from the toroidal to poloidal fields. The Faraday
rotation measures of a good number of background radio sources
behind the central region have also been observed and used for the
modelling the field structure near the central bar.

The highly polarized non-thermal radio filaments (e.g. \cite{ymc84},
\cite{lme99}, \cite{yhc04}, \cite{lnlk04}) have been detected within
$1^{\circ}$ from the Galactic center. They are almost perpendicular to
the Galactic plane, although some of the newly found filaments are not
(\cite{lnlk04}). After Faraday rotation correction, polarization
observations show that the magnetic fields are aligned along the
filaments (\cite{lme99}). These filaments are probably
radio-illuminated flux tubes, with a field strength of about 1~mG
(\cite{ms96}), which indicate poloidal magnetic fields within a few
hundred parsecs from the Galactic center. The diffuse radio emission
detected in an extent of about 400~pc (\cite{lbs+05}) implies a weak
pervasive field of tens of $\mu$G. However, this could be the
volume-averaged field strength in such a region. The observed ``double
helix'' nebula (\cite{mud06}) with an estimated field strength on
order of 100~$\mu$G reinforces the presence of strong poloidal
magnetic fields in the form of tubes merging from the rotating
circumnuclear gas disk near the Galactic center.

Polarized thermal dust emission has been detected in the ring-like
central molecular cloud zone of a size of a few hundred pc at sub-mm
wavelength (\cite{ncr+03}, \cite{cdd+03}), which indicates the
orientations of toroidal fields in the clouds parallel to the Galactic
plane.  The Zeeman splitting measurements the OH maser emission
(\cite{yrg+99}) or absorption (\cite{ug95}) in the central region give
a line-of-sight field strength of 0.1 to a few mG in the clouds. It is
possible that toroidal fields in the clouds are sheared from the
poloidal fields.

Recently, new near-infrared observations have been made of the stars
in a sky area of $2^{\circ}\times2^{\circ}$ around the Galactic center
(\cite{nht+10}). Differential polarization between the objects behind
the center and the objects in the front of the center shows that the
magnetic fields are toroidal near the Galactic plane and smoothly
transition to poloidal fields at latitudes of $|b|>0.4^{\circ}$.

Within a few hundred pc to a few kpc from the center, stellar and gas
distributions and the magnetic structure are all mysterious. There
probably is a bar. If so, the large-scale magnetic fields should be
closely related to the material structure but have not been observationally revealed yet. The RMs of background radio sources within $|l|<6^{\circ}$
of the Galactic center (\cite{rrs05}) are consistent with either the
large-scale magnetic fields of a bisymmetric spiral configuration or
the large-scale fields oriented along the Galactic bar (\cite{rrs08}).

\subsection{Magnetic fields in the Galactic halo}

Our edge-on Milky Way has a tenuous extended component beyond the
obvious disk, which is shown in radio emission and the space
distribution of stars and gas. We call it the halo or the thick disk,
which should probably be defined as the component outside a
given height from the Galactic plane. However, it is difficult at
present to define the scale radius and scale height for the extension
of such a halo. In reality, the local emission features are always
mixed with the weak background halo component.  Observationally the
halo is always visible at Galactic latitudes higher than a few
degrees.

Magnetic fields in the Galactic halo have been shown by the starlight
polarization, synchrotron emission, the polarized emission of dust
grains and clouds and the sky distribution of rotation measures of
extragalactic radio sources and pulsars. The measurements of starlight
polarization of a few thousand stars (\cite{mf70a}; \cite{hei00})
within a few kpc from the Sun show not only the field orientations
mostly parallel to the Galactic plane in the disk but also the fields
in the local halo of a few kpc (\cite{hei96}). The most prominent
feature is the polarization vectors following the north Galactic spur,
which in fact is a very localized feature.

The best evidence for the existence of magnetic fields in such a halo
is the radio continuum emission, much extended away from the disk
(\cite{bkb85}), as seen in the early days at 408 MHz (\cite{hssw82}) and
later at 1420 MHz (\cite{rr86, rtr01}). The polarized diffuse emission
at a few hundred MHz mainly comes from regions within a few hundred
parsecs from the Sun while the polarized emission at higher
frequencies from more distant regions. In the all sky polarization map
at 1.4 GHz (\cite{wlrw06,trr08}), strong polarization is seen in
the outer Galaxy and high Galatic latitutes. Obvious depolarization is
seen in the inner Galaxy near the Galactic plane. Recent development
of spectro-polarimetry technique has been applied for the diffuse halo
emission, and the sky has been mapped in many polarization channels
in a few hundred MHz (\cite{wfl+10}). Synthesized rotation measure
maps\footnote{Rotation measure synthesis is probably not a good tool
  to separate the diffuse emission from different regions along a
  sightline at different distances with different
  polarization properties and different foreground Faraday rotations.}
can be constructed. Some features emerged in some rotation measure
channels are coincident or anti-coincident with some known
objects. The best polarization maps which suffer less depolarization
come from the WMAP measurements (\cite{phk+07, jbd+11}), which show
the well-ordered distribution of E-vectors. In the disk near the
Galactic plane, the indicated magnetic fields are parallel to the
Galactic plane, while beyond the plane, the B vectors are indicators
of the local halo fields, showing the transition from the horizontal to
vertical field components. In other words, these vectors at hight
latitudes are very useful for constraining the local halo fields.

Because of the geometry of the Galactic halo and our location at the
disk edge, the Faraday rotation measures in the middle latitude
regions of the inner Galaxy ($|l|<90^{\circ}$) contain the information
of the magnetic field component along the line of sight and hence are
the best probes for the magnetic fields in the halo. The rotation
measure distribution as well as other polarization measurements in the
outer Galaxy (\cite{mmg+12}) cannot be used to show the halo fields
because the observed RMs of background sources contain various
intrinsic RMs and probably different RM contributions from
intergalactic space, in addition to the common RM foreground from our
Milky Way. The outliers of RM data have to be filtered out
and then the average or the median of the RM data can be the
representative for the RM foreground. The antisymmetry of the RM sky
so found by Han et al.  (1997)\nocite{hmbb97} in the inner Galaxy has
been interpreted as the toroidal magnetic fields in the halo with 
reversed field directions below and above the Galactic plane (see also
\cite{hmq99}), which are very consistent with the magnetic field
configuration of the A0 dynamo.  The vertical fields in the Galactic
center naturally act as the poloidal fields in such a field
configuration. Although the scale radius and scale height for such
halo toroidal fields are not yet known, and more seriously
little is known about the electron density distribution in the halo,
this qualitative toroidal halo field model derived from the
antisymmetric RM sky (\cite{hmbb97,hmq99}) has now widely been adopted
in many quantitative models (e.g. \cite{srwe08}) since the first
parametrization by \cite{ps03}.  As more and more RM data have become
available (\cite{tss09}), the antisymmetric RM sky has been 
confirmed, no matter how data are analyzed and presented (e.g.
\cite{ojr+12}). Certainly the much enhanced density of the RM data
distribution can show the details of magnetic fields (\cite{sts11})
especially in large objects (\cite{hmg11}), in addition to the
large-scale RM distribution for the halo fields.

Using the rotation measure distribution in the two Galactic poles, the
local vertical field component can be estimated, which could be taken
as a part of the halo field or the dipole field. In the early days,
using a few tens of RMs, we found $B_z \sim 0.2 \mu$G
(\cite{hq94,hmq99}).  Recent extensive observations have resulted in a
much improved RM dataset (\cite{mgh+10}) which more or less confirmed
the existence and the strength of such a vertical field.

\subsection{Magnetic fields in the Galactic disk}

The stellar disk of the Milky Way is very thin, only a few tens of pc
in thickness. The diffuse interstellar  medium in the disk has a 
larger thickness of at least a hundred pc. In the disk, there are
spiral arms which can be traced well by molecular clouds and HII regions
(\cite{hhs09}). But we do not know how many arms exist in our Milky
Way and how they are connected.

Magnetic fields in the Galactic disk have been measured by all five
probes (see discussion on advantages of these probes in
\cite{han13}). Some of these results have been mentioned above.  Three
probes, starlight polarization, polarized emission from dust grains
and clouds at millimeter and submillimeter wavelengths, and diffuse
radio synchrotron emission from relativistic electrons in interstellar
magnetic fields, give the magnetic field orientation projected onto
the sky plane. Because of our location in the Milky Way, they are not
very good probes for {\it large-scale magnetic structure} in the
Galactic disk.

Starlight polarization data near the Galactic plane show that the
orientation of magnetic fields in clouds, traced by the accumulated
polarization due to absorption of dust grains preferentially oriented
by the fields, are always parallel to the Galactic plane, except for
high Galactic latitude data for very local features (\cite{mf70a},
\cite{hei96}, \cite{nht+10}). Deep observations (\cite{cppt12}) will
probably reach the same conclusion with much more and new data. A
similar situation occurs for polarization measurements of thermal
emission from molecular clouds at mm, submm or infrared
wavelengths. Observed magnetic fields in most clouds are parallel to
the Galactic plane (\cite{ncr+03}, \cite{lgk+06}, \cite{bmd+11}), no
matter how far away or where the clouds are located in the disk,
except that the large-scale fields are distorted locally. From such
orientation measurements of cloud fields, it is impossible to derive
the large-scale magnetic fields in the disk. Nevertheless the results
suggest that the fields in clouds have their orientations well
preserved during cloud formation probably from the similarly oriented
magnetic fields frozen in very diffuse interstellar gas.

The diffuse synchrotron emission comes from everywhere in the
disk. The observed emission is very depolarized in the low latitudes
at low frequencies because the emission from different distances
suffers different Faraday rotations and then is added together for the
observed values. Therefore it is not possible to derive the
large-scale magnetic field structure in the disk from the observed
polarized synchrotron emission at low frequencies. At a frequency
of a few GHz, the depolarization occurs in the inner Galaxy
(\cite{srh+11,xhr+11}) but not the outer Galaxy (\cite{grh+10}) where
many magnetic field details can be seen around various objects, such
as SNRs and HII regions. At very high frequencies above 20~GHz, no
severe depolarization occurs even in the inner Galaxy.
Polarization measurements simply indicate that the magnetic fields are
parallel to the Galactic plane (\cite{phk+07}). It is not
possible to derive the large-scale magnetic field structure in the
disk from the polarization measurements of diffuse synchrotron
emission at low latitudes, though some models which include the
magnetic field structure can fit well the variations of total
synchrotron emission intensity along the Galactic longitude,
which result from the enhanced field strength in spiral arms.

Two kinds of measurements for the field component along the line of
sight can be used for the large-scale field structure. One is the
Zeeman splitting of line emission or absorption, the other is the
Faraday rotation of background sources and pulsars. The Zeeman
splitting can be used to measure the fields (both strength and
directioon) in clouds or clumps of a scale of pc or even
AU. Surprisingly, the distribution of available magnetic field
measurements from maser lines around a large number of HII regions is
very coherent with the large-scale spiral structure and large-scale
magnetic fields (\cite{hz07}). This is strong evidence that magnetic
fields have preserved their directions from the kpc scale to AU scale
during cloud formation and star formation processes. A big project for
extensive observations of many HII regions for the coherent
large-scale magnetic structure of the Milky Way is in progress
(\cite{gmc+12}).

Faraday rotation measures of background radio sources behind the
Galactic disk can be very powerful probes for the field structure.
Many authors have tried to figure out the disk field structure from
the RM variation as a function of Galactic longitude
(\cite{sk80,sf83,ptkn11}).  The median or average RMs of background
sources behind the disk are the integrated measurement of polarization
angle rotations over the whole path in the disk and therefore are not
sensitive to the possible magnetic field reversals inside the disk
between arms and interarm regions along the path. The dominant
contribution to RMs of background sources comes from tangential
regions of spiral arms, where the magnetic fields have the smallest
angle with the line of sight if the large-scale magnetic fields follow
spiral arms. In recent years extensive efforts have been made to
enlarge the RM samples of background sources at low Galactic latitudes
(e.g. \cite{bhg+07,vbs+11}). The RM data behind the Galactic disk (see
Fig.1 of \cite{han13}) are much denser in the outer Galaxy than in the
inner Galaxy. Closer to the Galactic center, the RM data become more
scarce, because the diffuse emission is stronger and because the
polarization observations are more difficult to carry out. The RM data
in the outer Galaxy cannot constrain the magnetic field structure in
the Galactic disk, because such large-scale fields become more and
more perpendicular to the line of sight. When a set of RMs of
background sources is fitted with a magnetic structure model, the
electron density model is a necessary input. The disk magnetic field
models derived mainly from the RMs of background sources along 
Galactic longitude should be treated with cautions because the field
reversals in the disk cannot be constrained.

Pulsars are excellent probes of the magnetic fields in our Milky
Way. They are located inside our Galaxy. The observed RMs of pulsars
come only from the interstellar medium between pulsars and us, because
there is no intrinsic Faraday rotation from the emission region and
pulsar magnetosphere. For a pulsar at distance $D$ (in
pc), the RM is given by
$ 
{\rm RM} = 0.810 \int_{0}^{D} n_e {\bf B} \cdot d{\bf l}.
$ 
With the pulsar dispersion measure,
$ 
{\rm DM}=\int_{0}^{D} n_e d l,
$ 
we obtain a direct estimate of the field strength weighted by the local free
electron density
\begin{equation}\label{eq_B}
\langle B_{||} \rangle  = \frac{\int_{0}^{D} n_e {\bf B} \cdot d{\bf
l} }{\int_{0}^{D} n_e d l } = 1.232 \;  \frac{\rm RM}{\rm DM}.
\label{eq-B}
\end{equation}
If pulsar RM data are model-fitted with the magnetic field
structures with the electron density model (\cite{hq94,id99,njkk08,nk10}),
then the pulsars and EGRs are more or less equivalently good as
probes for the magnetic structure. However, pulsars are spread
through the Galaxy at approximately known distances, allowing three-dimensional mapping of the magnetic
fields. When RM and DM data are available for many
pulsars in a given region with similar lines of sight, e.g. one pulsar
at $d0$ and one at $d1$, the RM change against distance or DM can
indicate the direction and magnitude of the large-scale field in
particular regions of the Galaxy (\cite{hmq99,hmlq02,hml+06,njkk08}). Field
strengths in the region can be directly derived by using
$ 
\langle B_{||}\rangle_{d1-d0} = 1.232 \frac{\Delta{\rm RM}}{\Delta{\rm DM}},
$ 
where $\langle B_{||}\rangle_{d1-d0}$ is the mean line-of-sight field
component in $\mu$G for the region between distances $d0$ and $d1$,
$\Delta{\rm RM} = {\rm RM}_{d1} - {\rm RM}_{d0}$ and $\Delta{\rm DM}
={\rm DM}_{d1} - {\rm DM}_{d0}$. Notice that this derived field is
not dependent on the electron density model.

Using pulsar data, Han et al. (2006)\nocite{hml+06} show that magnetic
fields in the spiral arms (i.e. the Norma arm, the Scutum and Crux
arm, and the Sagittarius and Carina arm) are always counterclockwise
in both the first and fourth quadrants, though some disordered fields
appear in some segments of some arms. At least in the local region and
in the fourth quadrant, there is good evidence that the fields in
interarm regions are similarly coherent, but reversed to be
clockwise. The strengths of regular azimuthal fields near the
tangential regions in the 1st and 4th Galactic quadrants show a clear
tendency that the fields get stronger at smaller Galactocentric
radius. It has been found from pulsar RMs that the random field has a
strength of $B_r\sim 4-6\; \mu$G independent of cell-size in the scale
range of 10 -- 100~pc (e.g. \cite{os93}). From pulsar RMs in a very
large region of the Galactic disk, Han et al. (2004)\nocite{hfm04}
obtained a power law distribution for magnetic field fluctuations of
$E_B(k)= C \ (k / {\rm kpc^{-1}})^{-0.37\pm0.10}$ at scales from
$1/k=$ 0.5~kpc to 15~kpc, with $C= (6.8\pm0.8)\ 10^{-13} {\rm erg
  \ cm^{-3} \ kpc}$, corresponding to an rms field of $\sim6\; \mu$G in
the scale range.

\section{Conclusions}

Magnetic fields in our Milky Way and nearby galaxies have been probed
by using five types of observations, each only giving partial
information of one of the 3D field components. Magnetic field
structure and field strength have been derived from these
measurements. The magnetic fields in spiral galaxies follow the spiral
structure and are passive to dynamics. Magnetic field directions are
evidently preserved during the cloud formation from diffuse interstellar
gas and the star formation regions.  Magnetic fields in our Milky Way
show field direction reversals in the halo and in the disk between
arms.  Such reversals have not yet been detected in nearby
galaxies. We know more details about magnetic fields in our Milky Way
than in nearby galaxies.  Obviously more sensitive observations are
needed for the field strength and field structure in nearby galaxies,
via diffuse polarization observations in multi-frequency channels,
Zeeman splitting observations of clouds and masers, and the polarized
thermal emission of clouds. Such observations will be possible in the
near future using ALMA and in distant future using the SKA.

\begin{acknowledgments}
I am grateful to Dr. Tim Robishaw who encouraged me to write down all
of my understanding of the topic and careful read the manuscript.  The
author is supported by the National Natural Science Foundation of
China (10833003).
\end{acknowledgments}

\end{document}